\documentclass[12pt]{article}
\pdfoutput=1
\usepackage{amssymb,amsmath}
\usepackage{graphicx}
\usepackage{subfig}
\usepackage{color}
\usepackage{slashed}
\usepackage{enumerate}
\usepackage{dsfont}
\usepackage{hyperref}
\definecolor{nicered}{rgb}{0.7,0.1,0.1}
\definecolor{nicegreen}{rgb}{0.1,0.5,0.1}
\hypersetup{colorlinks,citecolor= nicegreen,linkcolor= nicered}

\allowdisplaybreaks
\begin{document}
\begin{titlepage}

  \newcommand{\AddrJSI}{{\sl \small J. Stefan Institute, \\\sl \small
      Jamova 39, P. O. Box 3000, 1001 Ljubljana, Slovenia}}
  \newcommand{\AddrFMF}{{\sl \small Department of Physics, University
      of Ljubljana, \\\sl \small Jadranska 19, 1000 Ljubljana,
      Slovenia}} 
  \newcommand{\AddrLiege}{{\sl \small IFPA, Dep. AGO,
      Universite de Liege, \\ \small \sl Bat B5, Sart Tilman B-4000
      Liege 1, Belgium}} 
  \vspace*{1.5cm}
\begin{center}
  \textbf{\large Minimal lepton flavor violating realizations
    \\[2mm]
    of minimal seesaw models}
  \\[15mm]
  D. Aristizabal Sierra$^{a,}$\footnote{e-mail address:
    daristizabal@ulg.ac.be}, A. Degee$^{a,}$\footnote{e-mail address:
    audrey.degee@ulg.ac.be}, J. F. Kamenik$^{b,c,}$\footnote{e-mail address:
    jernej.kamenik@ijs.si} 
  \vspace{1cm}\\
  $^a$\AddrLiege.\vspace{0.4cm}\\
  $^b$\AddrJSI.\vspace{0.4cm}\\
  $^c$\AddrFMF.\vspace{0.4cm}\\
\end{center}
\vspace*{0.5cm}
\begin{abstract}
\footnotesize
  We study the implications of the global $U(1)_R$ symmetry present in
  minimal lepton flavor violating implementations of the seesaw mechanism for neutrino masses. In the
  context of minimal type I seesaw scenarios with a slightly broken $U(1)_R$, we
  show that, depending on the $R$-charge assignments, two classes of
  generic models can be identified. Models where the right-handed
  neutrino masses and the lepton number breaking scale are decoupled,
  and models where the parameters that slightly break the $U(1)_R$ induce
  a suppression in the light neutrino mass matrix. 
  We show that within the first class of models, contributions of right-handed neutrinos to charged lepton flavor violating
  processes are severely suppressed.  Within the second
  class of models we study the charged lepton flavor violating
  phenomenology in detail, focusing on $\mu\to e\gamma$, $\mu\to 3e$ and $\mu-e$
  conversion in nuclei. We show that sizable contributions to these processes are naturally obtained for right-handed neutrino masses at the TeV scale.
  We then discuss the interplay with the effects of the right-handed neutrino
  interactions on primordial $B-L$ asymmetries, finding
  that sizable right-handed neutrino contributions to charged lepton flavor violating processes are incompatible
  with the requirement of generating (or even preserving preexisting) $B-L$ asymmetries
  consistent with the observed baryon asymmetry of the Universe.
\end{abstract}
\end{titlepage}

\setcounter{footnote}{0}

\section{Introduction}
\label{sec:intro}
The observation of neutrino flavor oscillations constitutes an
experimental proof of lepton flavor violation
\cite{Schwetz:2011zk}. In principle, other manifestations of such
effects could be expected to show up in the charged lepton sector as
well. However, the lack of a definitive model for neutrino mass
generation implies that conclusive predictions for lepton flavor
violating processes can not be made, and even assuming a concrete
model realization for neutrino masses, predictions for such effects
can only be done if the flavor structure of the corresponding
realization is specified.

A problem that one faces when dealing with charged lepton flavor
violating phenomenology is related with the arbitrariness of the free
parameters that define these observables.  In this regards the minimal
lepton flavor violation (MLFV) hypothesis
\cite{Cirigliano:2005ck,Davidson:2006bd,Gavela:2009cd,Alonso:2011jd}
is a very useful guide for constructing predictive models in which
lepton violating signals are entirely determined by low-energy
neutrino data. The relationship between the free parameters and
neutrino observables can arise via either a restrictive MLFV
hypothesis or in some cases due to the intrinsic structure of the
corresponding model (whenever the number of free parameters is
comparable to the number of neutrino observables).

In the canonical seesaw (type-I) the kinetic sector of the model is
invariant under a global large flavor symmetry group
$G_F=SU(3)_e\times SU(3)_\ell\times SU(3)_N$, where $\ell$, $e$ and
$N$ denote triplets in flavor space constructed from the electroweak
lepton doublets and singlets and RH neutrinos. In addition there is an
invariance under an extra global $U(1)_R$
\cite{D'Ambrosio:2002ex,Alonso:2011jd}.  In principle the charges
associated with this global transformation (hereafter denoted by $R$)
are arbitrary, and thus different $R$-charge assignments define
different models with their own consequences for lepton flavor
violating phenomenology. In particular, models for which the $R$
charges allow for large Yukawa couplings and TeV RH neutrino masses
should lead to sizable charged lepton flavor violating processes.

Consistent models of ${\cal O}(\mbox{TeV})$ RH states and large Yukawa
couplings are achievable if cancellations among different pieces of
the light neutrino mass matrix are allowed, and the RH neutrino mass
spectrum is not strongly hierarchical~\cite{Ibarra:2010xw}\footnote{In
  the case of a large hierarchy among the different RH neutrino masses
  the one-loop finite corrections to the light neutrino mass matrix
  can exceed the corresponding tree-level contributions. Neglecting such corrections can in this case
   lead to a model inconsistent with neutrino data
 ~\cite{AristizabalSierra:2011mn}.}. The class of TeV scale seesaw
models arising from the presence of the $U(1)_R$ symmetry are expected
to be in that sense different: no cancellations are needed because the
suppression in the effective light neutrino mass matrix is no longer
constrained to be related with the heaviness of the RH neutrinos.

In this paper we study the implications of a slightly broken $U(1)_R$
symmetry in minimal type-I seesaw models (with two RH neutrinos)
\footnote{Models with an arbitrary number of RH neutrinos were
  considered for the first time in \cite{Schechter:1980gr}.}
\footnote{For related discussion within Type III and mixed Type I+III
  seesaw scenarios c.f.~\cite{Kamenik:2009cb}.}.  We show that the
intrinsic structure of the relevant models leads to a flavor pattern
completely determined by low-energy neutrino observables, thus
realizing in that way the MLFV hypothesis. This feature in addition to
a slightly broken $U(1)_R$ leads to sizable $\mu\to e\gamma$, $\mu\to
3e$ and $\mu-e$ processes.  This is in contrast to models where
$U(1)_L$ is slightly broken so that lepton number and lepton flavor
violation occur at the same scale.

In models with slightly broken lepton number it has been shown that
leptogenesis is reconcilable with large charged lepton flavor
violating rates, as the washouts induced by the RH neutrino states are
controlled by the amount of lepton number violation~\cite{Blanchet:2009kk}. We show however, that in the class of models
considered here this statement does not hold, and that indeed
leptogenesis and  sizable charged lepton flavor violating effects are
mutually exclusive. However, since both phenomena cover non-overlapped
regions of parameter space their analyses are complementary. We
therefore also explore the constraints on these models derived from the
requirement of not erasing---via the RH neutrino dynamics---a
preexisting $B-L$ asymmetry below the value consistent with the
observed baryon asymmetry of the Universe.

The rest of this paper is organized as follows: in
section~\ref{sec:review} we discuss in detail the scenarios arising
from the different $R$-charge assignments. In
section~\ref{sec:lfv-processes} we discuss the phenomenology of
charged lepton flavor violating decays in the representative models.
In section~\ref{sec:primordial-asymm} we analyze the implications of such constructions for scenarios of high scale baryogenesis by
quantifying---via the Boltzmann equations---the effects of the RH neutrino
dynamics on preexisting $B-L$ asymmetries. In section~\ref{sec:conc}
we present our conclusions and final remarks.  Explicit formulas used
in the calculation of the different lepton flavor violating processes
under study are given in
appendix~\ref{sec:formulas-for-lfvprocesses}.
\section{The setups}
\label{sec:review}
The kinetic and gauge interaction Lagrangian of the standard model extended with two RH
neutrinos exhibits a global $G=U(3)_e\times U(3)_\ell\times U(2)_N$
symmetry. Factorizing three $U(1)$ factors from $G$, the global
symmetry can be rewritten as $U(1)_Y\times U(1)_L\times U(1)_R\times
G_F$ where $U(1)_{Y,L}$ can be identified with global hypercharge and
lepton number whereas the $U(1)_R$ is a ``new'' global
symmetry~\cite{D'Ambrosio:2002ex,Alonso:2011jd}. The remaining direct
product group $G_F=SU(3)_e\times SU(3)_\ell\times SU(2)_N$ determines
the flavor symmetry which is explicitly broken in the Yukawa sector.

In minimal lepton flavor violating seesaw models the Yukawa (mass)
matrices are treated as spurion fields transforming under $G_F$ in
such a way that the corresponding Yukawa (mass) terms in the leptonic
Lagrangian remain invariant under the global flavor symmetry. The
usual procedure is then based on an effective theory approach in which
a set of non-renormalizable effective operators are constructed from
the spurions.\footnote{An exception are the explicit MLFV models
  discussed in Ref.~\cite{Gavela:2009cd}.} With the operators at hand
and under certain restrictions on $G_F$ the lepton flavor violating
effects can be estimated by means of low-energy neutrino
data~\cite{Cirigliano:2005ck,Alonso:2011jd}. Here, instead, we
explicitly consider the seesaw Lagrangian with a slightly broken
$U(1)_R$ and classify the possible realizations according to the
$R$-charge assignments. Under this consideration, in the models
featuring sizable lepton flavor violating effects, the flavor
structure is determined by low-energy observables as well (up to
a global normalization factor), not as a consequence of a restricted
MLFV hypothesis but by the intrinsic structure of the resulting models.

Depending on the $R$-charge assignments two classes of generic models
can be identified. Let us discuss this in more detail. Requiring
$U(1)_R$ invariance of the charged lepton Yukawa terms determines
$R(e)$ in terms of $R(\ell,H)$. After fixing $R(H)=0$, to avoid
charging the quark sector, the remaining charges can be fixed by
starting with $R(N_{1,2})$. In order to have sizable lepton flavor
violating effects both the $N_{1,2}$ Majorana mass terms should be
suppressed by $R$-breaking parameters (generically denoted by
$\epsilon$), so $R(N_{1,2})\neq 0$. Thus one has only three
possibilities: (A) $R(N_1)=R(N_2)$, (B) $R(N_1)=-R(N_2)$ and (C)
$|R(N_1)|\neq |R(N_2)|$. The phenomenology of case (C), however, is
expected to be similar to the one arising from models with $R$-charge
assignments of type (A). The reason being that in that case the
$N_1-N_2$ mixing is always $U(1)_R$ suppressed, which in turn implies
suppressed charged lepton flavor violating processes (equivalently,
the effective neutrino mass matrix will not be $\epsilon$ suppressed,
forcing tiny Yukawas or heavy $N_{1,2}$ states). This is to be
compared with models based on $R$-charge assignments of type (B),
where the $N_1-N_2$ turns out to be maximal and a set of unsuppressed
$\ell-N$ Yukawa couplings can be always obtained by properly chosing
$R(\ell)$ (in these models the effective neutrino mass matrix involves
extra $\epsilon$ suppression factors).

In summary depending on the $R$-charge assignments two classes of
generic models can be identified: models in which the mechanism that
suppresses the light neutrino masses propagates to lepton flavor
violating observables, thus rendering their values far below planned
experimental sensitivities; and models in which the mechanism
decouples in such a way that lepton flavor violating effects become
sizable.  In what follows, in order to illustrate this is actually the
case, we will discuss two examples of models of type A and B.
~

\noindent
\underline{\bf Type A models}~\cite{Alonso:2011jd}: $R(N_a)=+1$ and
$R(\ell_i,e_i,H)=0$
($H$ being the Higgs electroweak doublet)\\
In this case, in the basis in which the charged lepton Yukawa
couplings and the Majorana RH neutrino mass matrix are diagonal, the
Lagrangian reads
\begin{equation}
  \label{eq:seesaw-lag}
  {\cal L}=
  -\bar \ell\,\pmb{\hat Y_e}\,e H 
  - \epsilon\, \bar \ell\,\pmb{\lambda}^*\,N \tilde H
  - \frac{1}{2} \epsilon^2\,\mu N^T\,C\,\pmb{\hat Y_N}\,N + \mbox{h.c.}\,.
\end{equation}
Here $\tilde H = i\sigma_2 H^*$, $C$ is the charge conjugation
operator and $\pmb{\hat Y_e}$, $\pmb{\lambda}^*$ and $\pmb{\hat Y_N}={\rm diag}(Y_{N_1},Y_{N_2})$
are the Yukawa coupling matrices (we denote matrices in
bold-face). The dimensionless parameter $\epsilon\ll 1$ slightly
breaks $U(1)_R$ whereas, due to the assignment $L(N)=1$, the lepton
number $U(1)_L$ factor is broken by $\mu$. With this setup the
$5\times 5$ neutral fermion mass matrix can be written as
\begin{equation}
  \label{eq:neutral-fermion-mm}
  \pmb{M_N}=
  \begin{pmatrix}
    \pmb{0} & \epsilon\,v\pmb{\lambda}\\
    \epsilon\,v\pmb{\lambda}^T & \epsilon^2\,\mu\pmb{\hat Y_N}
  \end{pmatrix}\,,
\end{equation}
where $\langle H \rangle = v\simeq 174$ GeV. In the seesaw limit,
which in this case reads
$v\pmb{\lambda}\ll\epsilon\mu\pmb{\hat Y_N}$ the effective
light neutrino mass matrix is given by
\begin{equation}
  \label{eq:eff-nmm}
  \pmb{m_\nu^\text{eff}}=-\frac{v^2}{\mu}\sum_{a=1,2}
  \frac{\pmb{\lambda_a}\otimes\pmb{\lambda_a}}{Y_{N_a}}\,,
\end{equation}
with $\pmb{\lambda_a}=(\lambda_{e a},\lambda_{\mu a},\lambda_{\tau a})$.
The corresponding light neutrino masses are obtained from the
leptonic mixing matrix $\pmb{U}=\pmb{V}\,\pmb{\hat P}$ (with $\pmb{V}$
having a CKM form and $\pmb{\hat
  P}=\mbox{diag}(e^{i\phi},e^{i\phi},1)$ containing the Majorana
CP phase) after diagonalization:
\begin{equation}
  \label{eq:diagonalization}
  \pmb{U}^T\,\pmb{m_\nu^\text{eff}}\,\pmb{U}=
  \pmb{\hat m_\nu^\text{eff}}\,.
\end{equation}
In the 2 RH neutrino mass model the constrained parameter space
enforces one of the light neutrinos to be massless. Thus, in the
normal hierarchical mass spectrum case $m_{\nu_1}=0$ and
$m_{\nu_2}<m_{\nu_3}$ whereas in the inverted case $m_{\nu_3}=0$ and
$m_{\nu_1}<m_{\nu_2}$.

Since the dimension five effective operator is $U(1)_R$ invariant the
neutrino mass matrix does not depend on $\epsilon$; the suppression
ensuring light neutrino masses is solely provided by the lepton number
breaking parameter $\mu$. On the other hand, the RH neutrino mass spectrum is determined
by
\begin{equation}
  \label{eq:RHneutrino-masses}
  \pmb{\hat {\cal M}_N}=\epsilon^2\,\mu\,\pmb{\hat Y_N}\,.
\end{equation}
From this expression it can be seen that as long as the $U(1)_R$
global symmetry is an approximate symmetry of the Lagrangian
($\epsilon\ll 1$) the RH neutrino mass scale is decoupled from the
lepton number violating scale. Thus, the RH neutrino masses do not lie
at the same scale at which lepton number breaking takes place.

Assuming $\pmb{\hat Y_N},\pmb{ \lambda}\lesssim {\cal O}(1)$ an estimation of  the lepton number breaking parameter
$\mu\sim 10^{15}$ GeV can be obtained using $\sqrt{\Delta m^2_{31}} \sim 0.05$~eV~\cite{PDG} as a measure of the largest light neutrino mass in these scenarios.  From this estimation
and eq.~(\ref{eq:RHneutrino-masses}) it can be seen that values of
$\epsilon$ of the order of $\sim 10^{-6}$ allow to lower the lightest
RH neutrino mass below 1 TeV.

Formal invariance of the Lagrangian under $G_F$ is guaranteed if the
Yukawa matrices, promoted to spurion fields, transform according to
\begin{equation}
  \label{eq:spurions}
  \pmb{Y_e}\sim(\pmb{\bar 3}_e,\pmb{3}_\ell,\pmb{1}_N)\,,\quad
  \pmb{\lambda}^*\sim(\pmb{1}_e,\pmb{3}_\ell,\pmb{\bar 2}_N)\,,\quad
  \pmb{Y_N}\sim(\pmb{1}_e,\pmb{1}_\ell,\pmb{\bar 3}_N)\,.
\end{equation}
The constraints imposed by $G_F$ imply
\begin{equation}
  \label{eq:lambda-models-typeA}
  \pmb{\lambda}=\pmb{\lambda_\ell}\otimes\pmb{\lambda_N}\,,
\end{equation}
where $\pmb{\lambda_\ell}$ is a $SU(3)_\ell$ triplet and
$\pmb{\lambda_N}$ a $SU(2)_N$ doublet in flavor space. Accordingly, in
these kind of models a unequivocal determination of the flavor
structure via the MLFV hypothesis is possible by means of a
restrictive flavor symmetry $G_F'\subset G_F$. Though several
possibilities may be envisaged we do not discuss further details since, as
we show below, in this type of models contributions to lepton flavor violating processes of charged leptons are always negligible.

~

\noindent
\underline{\bf Type B models}: $R(N_1,\ell_i,e_i)=+1$, $R(N_2)=-1$,
$R(H)=0$.  Changing $R$ charges to $L$ charges, this case resembles
models where lepton number is slightly broken (see for example
\cite{Gavela:2009cd,Mohapatra:1986bd,Gu:2008yj,Ibanez:2009du}). The
Lagrangian is given by
\begin{equation}
  \label{eq:seesaw-lag-N!N2mismatch}
  {\cal L}=
  -\bar \ell\,\pmb{\hat Y_e}\,e H
  - \bar \ell\,\pmb{\lambda_1}^*\,N_1 \tilde H
  - \epsilon_\lambda\, \bar \ell\,\pmb{\lambda_2}^*\,N_2 \tilde H
  - \frac{1}{2} N_1^T\,C M\,N_2
  - \frac{1}{2}\epsilon_N N_a^T\,C M_{aa}\,N_a + 
  \mbox{h.c.}\,.
\end{equation}
The $\epsilon_{\lambda,N}$ parameters determine the amount of $U(1)_R$
breaking and are thus tiny. The diagonalization of the Majorana RH
neutrino mass matrix leads to two quasi-degenerate states with masses
given by
\begin{equation}
  \label{eq:RHN-masses-quasideg}
  M_{N_{1,2}}=M\mp\frac{M_{11}+M_{22}}{2}\epsilon_N\,.
\end{equation}
In the basis in which the RH neutrino mass matrix is diagonal the
Yukawa couplings become
\begin{equation}
  \label{eq:Yukawa-rotated}
  \lambda_{ka}\to -\frac{(i)^a}{\sqrt{2}}\left[\lambda_{k1}
    + (-1)^a\epsilon_\lambda\lambda_{k2}\right]\,,
  \qquad (k=e,\mu,\tau\;\;\mbox{and}\;\; a=1,2)\,,
\end{equation}
and the $5\times 5$ neutral fermion mass matrix is similar
as in type A models. However, due to the structure of the Yukawa
couplings the effective light neutrino matrix, up to ${\cal
  O}(\epsilon_N\epsilon_\lambda^2)$, has the following form
\begin{equation}
  \label{eq:light-nmm-N1N2mismatch}
  \pmb{m_\nu^\text{eff}}=-\frac{v^2\epsilon_\lambda}{M}
  |\pmb{\lambda_1}||\pmb{\Lambda}|
  \left(
    \pmb{\hat \lambda_1}^*\otimes\pmb{\hat \Lambda}^* 
    +
    \pmb{\hat \Lambda}^*\otimes\pmb{\hat \lambda_1}^*
  \right)\,,
\end{equation}
with
\begin{equation}
  \label{eq:Lambda}
  \pmb{\hat \Lambda}^*=\pmb{\hat \lambda_2}^*-
  \frac{M_{11}+M_{22}}{4M}\frac{\epsilon_\lambda}{\epsilon_N}
  \pmb{\hat \lambda_1}^*\,.
\end{equation}
Here with the purpose of relating the flavor structure of these models
with low energy observables, and following ref.~\cite{Gavela:2009cd}, we expressed the parameter space vectors
$\pmb{\lambda_1}, \pmb{\Lambda}$ in the light neutrino mass matrix in
terms of their moduli $|\pmb{\lambda_1}|, |\pmb{\Lambda}|$ and unitary
vectors $\pmb{\hat \lambda_1}, \pmb{\hat \Lambda}$. 

Note that in these models lepton number is broken even when $U(1)_R$
is an exact symmetry of the Lagrangian. However due to the Yukawa
structure and degeneracy of the RH neutrino mass spectrum at this
stage $\pmb{m_\nu^\text{eff}}=0$. Although a non-zero Majorana
neutrino mass matrix arises only once the $R$ breaking terms are
present this does not imply that in the absence of lepton number
violating interactions a Majorana mass matrix can be built. In that
case---as can seen from eq.~(\ref{eq:seesaw-lag-N!N2mismatch})---only
Dirac masses can be generated, as it must be.

Since $\epsilon_\lambda\ll 1$ small neutrino masses do not require
heavy RH neutrinos or small Yukawa couplings, thus potentially
implying large lepton flavor violating effects. In that sense, as
already stressed, these models resemble those in which lepton number
is slightly broken but with lepton number as well as lepton flavor
violation taking place at the same scale $M$.

In contrast to models of type A, in this case due to the structure of
the light Majorana neutrino mass matrix the vectors $\pmb{\lambda_1}$
and $\pmb{\Lambda}$ can be entirely determined by means of the solar
and atmospheric mass scales and mixing angles, up to the factors
$|\pmb{\lambda_1}|$ and $|\pmb{\Lambda}|$, without invoking a
restrictive MLFV hypothesis. The relations are different for normal
and inverted light neutrino mass spectra~\cite{Gavela:2009cd}:
\begin{itemize}
\item[$\bullet$] \underline{Normal hierarchical mass spectrum}
  \begin{align}
    \label{eq:Yukawas-normalS1}
    \pmb{\lambda_1}&=|\pmb{\lambda_1}|\;\pmb{\hat \lambda_1}=
    \frac{|\pmb{\lambda_1}|}{\sqrt{2}}
    \left(
      \sqrt{1+\rho}\,\pmb{U_3}^* + \sqrt{1-\rho}\,\pmb{U_2}^*
    \right)\,,\\
    \label{eq:Yukawas-normalS2}
    \pmb{\Lambda}&=|\pmb{\Lambda}|\;\pmb{\hat \Lambda}=
    \frac{|\pmb{\Lambda}|}{\sqrt{2}}
    \left(
      \sqrt{1+\rho}\,\pmb{U_3}^* - \sqrt{1-\rho}\,\pmb{U_2}^*
    \right)\,,
  \end{align}
where $\pmb{U_i}$ denote the columns of the leptonic mixing matrix and
\begin{equation}
  \label{eq:rho-and-r-NS}
  \rho=\frac{\sqrt{1+r}-\sqrt{r}}{\sqrt{1+r}+\sqrt{r}}\,,\qquad
  r=\frac{m_{\nu_2}^2}{m_{\nu_3}^2-m_{\nu_2}^2}\,.
\end{equation}
\item[$\bullet$] \underline{Inverted hierarchical mass spectrum}
\end{itemize}
\begin{align}
    \label{eq:Yukawas-invertedS1}
    \pmb{\lambda_1}&=|\pmb{\lambda_1}|\;\pmb{\hat \lambda_1}=
    \frac{|\pmb{\lambda_1}|}{\sqrt{2}}
    \left(
      \sqrt{1+\rho}\,\pmb{U_2}^* + \sqrt{1-\rho}\,\pmb{U_1}^*
    \right)\,,\\
    \label{eq:Yukawas-invertedS2}
    \pmb{\Lambda}&=|\pmb{\Lambda}|\;\pmb{\hat \Lambda}=
    \frac{|\pmb{\Lambda}|}{\sqrt{2}}
    \left(
      \sqrt{1+\rho}\,\pmb{U_2}^* - \sqrt{1-\rho}\,\pmb{U_1}^*
    \right)\,,
  \end{align}
with
\begin{equation}
  \label{eq:rho-and-r-IS}
  \rho=\frac{\sqrt{1+r}-1}{\sqrt{1+r}+1}\,,\qquad
  r=\frac{m_{\nu_2}^2-m_{\nu_1}^2}{m_{\nu_1}^2}\,.
\end{equation}
With these results at hand we are now in a position to calculate the
most relevant lepton flavor violating processes, which we discuss in
turn.

\section{Lepton flavor violating  processes}
\label{sec:lfv-processes}
In type A models the RH neutrino masses can be readily at the TeV
scale for $\epsilon\sim 10^{-6}$.  Since the Yukawa couplings scale
with $\epsilon$ as well, type A models are---in that sense---on the
same footing as the canonical type-I seesaw model i.e. TeV RH neutrino
masses imply tiny Yukawa couplings and thus negligible charged lepton
flavor violating decay branching ratios. The main difference is that in the canonical case, sizable charged lepton flavor violation can still be induced via fine-tuned cancelations in the effective neutrino mass matrix, while no such effects are possible in the minimal type A models, since the relevant couplings are completely determined by light neutrino mass and mixing parameters.

Type B models, in contrast,
may exhibit naturally large Yukawa couplings even for TeV scale RH
neutrino masses (or even lighter masses, depending on the value of the
$U(1)_R$ breaking parameter $\epsilon_\lambda$).  Accordingly, several charged lepton flavor violating
transition rates---induced by the RH neutrinos at the 1-loop
level---can in principle reach observable levels. In what follows we study the allowed mass and
Yukawa normalization factor ranges by considering the following lepton flavor violating
processes: $l_i\to l_j\gamma$, $l_i\to 3l_j$ and $\mu-e$ conversion in
nuclei.
\subsection{$l_i\to l_j\gamma$ processes}
\label{sec:radiative-decays}
Among these lepton flavor violating processes, presently
the $\mu\to e\gamma$ transition is most severely constrained. The MEG collaboration recently established
an upper bound of $2.4\times 10^{-12}$ at the 90\%
C.L.~\cite{Adam:2011ch}. For $\tau\to e\gamma$ and $\tau\to \mu\gamma$
on the other hand, the bounds are $3.3\times 10^{-8}$ and $4.4\times 10^{-8}$ at 90\%
C.L.~\cite{:2009tk}, respectively.  In the limit $m_{l_j}\ll m_{l_i}$
the partial decay width for $l_i\to l_j\gamma$ processes reads
\cite{Ilakovac:1994kj}
\begin{equation}
  \label{eq:ltolpgamma}
  \Gamma(l_i\to l_j\gamma)=\frac{\alpha\,\alpha_W^2}{256\,\pi^2}
  \,\frac{m_i^5}{M_W^4}\,
  \left|
    G_\gamma^{l_il_j}
  \right|^2
  =
  \frac{\alpha}{1024\,\pi^4}
  \,\frac{m_i^5}{M_W^4}\,
  \left|
    \left(
      \pmb{\lambda}\,\pmb{G_\gamma}\,\pmb{\lambda}^\dagger
    \right)_{ij}
  \right|^2\,,
\end{equation}
where $M_W$ is the $W^\pm$ mass, $\alpha_W=g^2/4\pi$ and the elements of
the diagonal matrix $\pmb{G_\gamma}$ are given in eq.~(\ref{eq:gfac})
in the appendix. This function is such that in the limit 
$M_{N_a}\gg M_W$, $(G_\gamma)_{aa} \to M^2_{W}/2 M_{N_a}^2$.  The corresponding decay branching
ratios are determined from the partial decay width after normalizing
to $\Gamma_\text{Tot}^{l_i}=\hbar\,\tau_{l_i}$, with $\tau_{l_i}$ the
$l_i$ charged lepton mean lifetime.
In the limit $r_a\ll 1$, using eq.~(\ref{eq:RHneutrino-masses}) and
taking into account the Yukawa rescaling $\pmb{\lambda}\to
\epsilon\pmb{\lambda}$ the decay branching ratio in type A models can
be written as
\begin{equation}
  \label{eq:BRtypeAmodels}
  \text{BR}(l_i\to l_j\gamma)\simeq
  \frac{\alpha}{4096\pi^4}
  \frac{m_i^5}{\mu^4\epsilon^4}
  \frac{1}{\Gamma_\text{Tot}^{l_i}}
  \left|
    \left(
      \pmb{\lambda}\,\pmb{Y_N}^{-2}\pmb{\lambda}^\dagger
    \right)_{ij}
  \right|^2\,.
\end{equation}
Thus, assuming ${\cal O}(\pmb{\lambda},\pmb{Y_N})\sim 1$, for which
$\mu\sim 10^{15}$ GeV and taking $\epsilon=10^{-6}$, the value
required for ${\cal O}(\text{TeV})$ RH neutrino masses, we get
$\text{BR}(\mu\to e\gamma)\simeq 10^{-30}$. This behavior, being
extensible to other lepton flavor violating processes, shows that in
type A models lepton flavor violating effects are negligibly small.

In type B models in contrast such lepton flavor violating effects may be sizable.  Using
expression~(\ref{eq:Yukawa-rotated}) for the Yukawa couplings,
neglecting the piece proportional to $\epsilon_\lambda$ and taking the
limit $M_{N_a}\gg M_W$ the decay branching ratios can be expressed in
terms of the parameters $\pmb{\lambda_1}$:
\begin{equation}
  \label{eq:4}
  \text{BR}(l_i \to l_j \gamma)\simeq\frac{\alpha}{1024 \pi^4}
  \frac{m_i^5}{M^4}\frac{|\pmb{\lambda_1}|^4}{\Gamma_\text{Tot}^{l_i}}
  \left|
    \hat\lambda_{i1}\;\hat\lambda_{j1}^*
  \right|^2\,.
\end{equation}
Since the components of the unitary vector $\pmb{\hat \lambda_1}$ are
entirely determined by low-energy observables (see
eqs. (\ref{eq:Yukawas-normalS1}) and (\ref{eq:Yukawas-invertedS1}))
the size of these branching ratios---and all the others discussed
below---are controlled only by the parameters $M$ and
$|\pmb{\lambda_1}|$, thus implying that for sufficiently light RH
neutrino masses and large $|\pmb{\lambda_1}|$ these processes may be
measurable.

In order to quantify the size of these lepton flavor violating effects we randomly
generate neutrino masses, mixing angles and Dirac and Majorana CP
violating phases in their $2\sigma$ ranges for both normal and
inverted hierarchical light neutrino mass
spectra~\cite{Schwetz:2011zk}. We also randomly generate the
parameters $|\pmb{\lambda_1}|$ and $M$ in the ranges
$[10^{-5},1]$ and $[10^2,10^6]$ GeV allowing RH neutrino
mass splittings in the range $[10^{-8},10^{-6}]$ GeV.  With the
numerical output we calculate the different lepton flavor violating decay branching ratios
from eq.~(\ref{eq:ltolpgamma}), using the full loop function given in
the appendix, eq.~(\ref{eq:gfac}). 
\begin{figure}
  \centering
  \includegraphics[width=7cm]{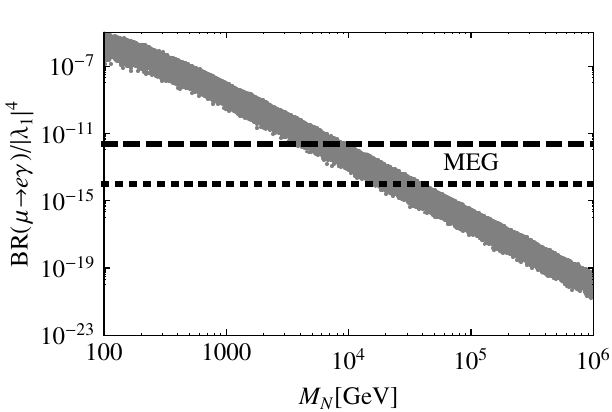}
  \includegraphics[width=7cm]{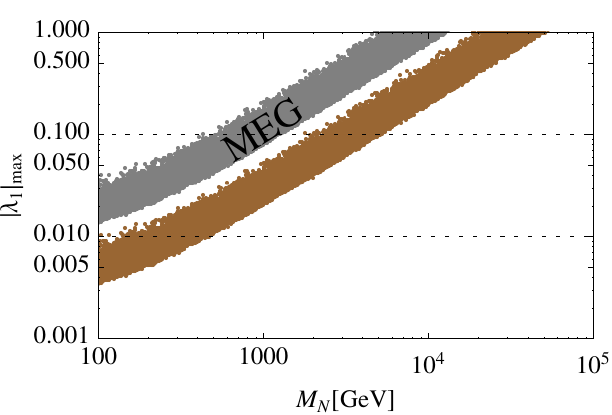}
  \caption{\it Decay branching ratio BR$(\mu\to e\gamma)$ normalized to $|\pmb{\lambda_1}|^4$ for
    normal 
    light neutrino mass spectrum as a function of the common RH
    neutrino mass (left hand side plot). The upper horizontal dashed line indicates the
    current limit on BR$(\mu\to e\gamma)$ from the MEG experiment
   ~\cite{Adam:2011ch}, whereas the lower dotted one marks prospective
    future experimental sensitivities~\cite{meg-futureS}. 
   The corresponding bounds on $|\pmb{\lambda_1}|$ from the present~\cite{Adam:2011ch} (upper gray band) and prospective future~\cite{meg-futureS} (lower brown band) experimental searches are shown in the right hand side plot. 
   The widths of the bands are due to the uncertainties in the neutrino mass matrix parameters.
    The results
    for the inverted neutrino spectrum are very similar and are thus
    not shown separately.}
  \label{fig:radiative-lfv}
\end{figure}

We find that radiative $\tau$ decay rates are always below their current bounds and
barely reach values of $10^{-9}$ for RH neutrino masses around 100-200
GeV (values exceeding the current bounds are not consistent with the
seesaw condition, that for concreteness we take as
$\pmb{m_D}\pmb{M_N}^{-1}<10^{-1}$), we thus focus on the $\mu\to
e\gamma$ process.  
The results for the normal mass spectrum case are
displayed in fig.~\ref{fig:radiative-lfv} as a function of the common
RH neutrino mass, $M_N=M$.
We observe that BR$(\mu\to e\gamma)$ can
reach the current experimental limit reported by the MEG experiment
\cite{Adam:2011ch} for RH neutrino masses $M_N<0.1~\mathrm{TeV},1~\mathrm{TeV},~10$~TeV provided $|{\pmb{\lambda_1}}|\gtrsim 2\times 10^{-2},~10^{-1}, ~1$, respectively.
The results for the inverted light neutrino mass hierarchy are very similar and consistent with these values.
Finally we note that the widths of the bands in fig.~\ref{fig:radiative-lfv} (and similarly for all the other considered processes below) are solely due to the uncertainties in the light neutrino mass matrix parameters (mainly $\theta_{13}$ and the CP violating phases) and can thus be improved with more precise light neutrino data.

\subsection{$l_i^-\to l_j^- l_j^- l_j^+$ processes}
\label{sec:lto3lp}
The decay branching ratios for these processes have been
calculated in~\cite{Ilakovac:1994kj}. The most constrained process in
this case is $\mu^-\to e^+e^-e^-$ for which the SINDRUM experiment has
placed a bound on the decay branching ratio of $10^{-12}$ at 90\%
C.L.~\cite{Bellgardt:1987du}. For $\tau^-\to e^+e^-e^-$ and $\tau^-\to
\mu^+\mu^-\mu^-$ the current bounds are $2.7\times 10^{-8}$ and
$2.1\times 10^{-8}$, respectively~\cite{Hayasaka:2010np}. The decay
branching ratios for these lepton flavor violating reactions are given
by~\cite{Ilakovac:1994kj}
\begin{align}
  \label{eq:brmuto3e}
  \mbox{BR}(l_i^-\to l_j^+l_j^-l_j^+)&=\frac{\alpha_W^4}{24576\pi^3}
  \frac{m_i^5}{M_W^4}
  \frac{1}{\Gamma^{l_i}_\text{Total}}
  \nonumber\\
  &\left\{2
    \left|
      \frac{1}{2}F_{\text{Box}}^{l_i3l_j} + F_Z^{l_il_j} - 2 s^2_W
      \left(
        F_Z^{l_il_j} - F_\gamma^{l_il_j}
      \right)
    \right|^2\right .
  \nonumber\\
  &\left .
    + 4s^4_W
    \left|
      F_Z^{l_il_j} - F_\gamma^{l_il_j}
    \right|^2
    +16s^2_W\mbox{Re}
    \left[
      \left(
        F_Z^{l_il_j} + \frac{1}{2}F_{\text{Box}}^{l_i3l_j}
      \right)G_\gamma^{l_il_j*}
    \right]
  \right .
  \nonumber\\
  &\left .
   -48 s_W^4\mbox{Re}
   \left[
     \left(
       F_Z^{l_il_j} - F_\gamma^{l_il_j}
     \right)G_\gamma^{l_il_j*}
   \right]
   + 32s_W^4\left|G_\gamma^{l_il_j}\right|^2
   \left(
     \ln\frac{m_i^2}{m_j^2} -\frac{11}{4}
   \right)
 \right\}\,.
\end{align}
Here $s_W=\sin\theta_W$, where $\theta_W$ is the weak mixing angle, and
the functions $F_\gamma^{l_i l_j}$, $F_Z^{l_i l_j}$ and $F^{l_i
  3l_j}_\text{Box}$ are form factors that involve the Yukawa couplings
and loop functions arising from the $\gamma$, $Z$ penguins and box
diagrams that determine the the full process (see appendix
\ref{sec:formulas-for-lfvprocesses} for a compilation of these
expressions and ref.~\cite{Ilakovac:1994kj} for their derivation).

\begin{figure}
  \centering
  \includegraphics[width=7cm]{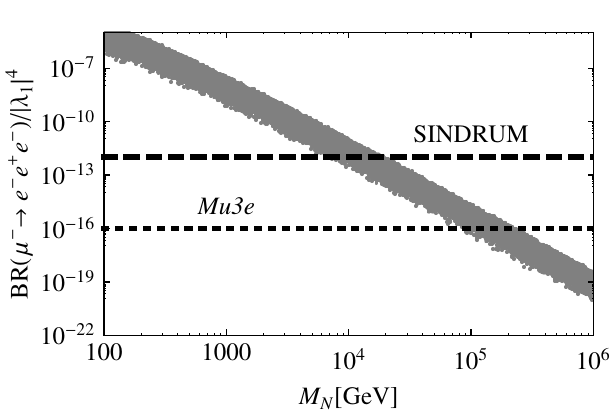}
  \includegraphics[width=7cm]{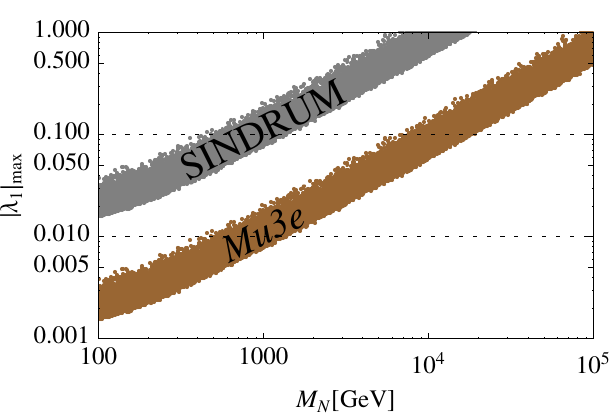}
  \caption{\it Decay branching ratio BR($\mu^-\to e^- e^+ e^-)$ normalized to $|\pmb{\lambda_1}|^4$
    for normal light neutrino
    mass spectrum as a function of common RH neutrino mass (left hand side plot). The
    upper horizontal dashed line indicates the current bound on the
    $\mu^-\to e^+e^-e^-$ rate placed by the SINDRUM experiment
   ~\cite{Bellgardt:1987du}, whereas the lower dotted one illustrates prospective future experimental
    sensitivities of the {\it Mu3e} experiment~\cite{psi}. 
    The corresponding bounds on $|\pmb{\lambda_1}|$ from the present~\cite{Bellgardt:1987du} (upper gray band) and prospective future~\cite{psi} (lower brown band) experimental searches are shown in the right hand side plot. 
   The widths of the bands are due to the uncertainties in the neutrino mass matrix parameters.    
    The results for the inverted neutrino spectrum are very similar and are thus not shown separately.}
  \label{fig:muto3elec}
\end{figure}

Following the same numerical procedure as in the $l_i\to l_j\gamma$
case and using the form factors given in the appendix we
evaluate the $\mu^+\to e^+e^-e^-$, $\tau^+\to e^+e^-e^-$ and
$\tau^+\to \mu^+\mu^-\mu^-$ decay branching ratios for both, the
normal and inverted light neutrino mass spectra. We find that
$\tau^+\to e^+e^-e^-$ and $\tau^+\to \mu^+\mu^-\mu^-$ processes are
always below $\sim 10^{-9}$ (due to the constraint enforced by the seesaw
condition when $|\pmb{\lambda_1}|>10^{-1}$), so in fig.~\ref{fig:muto3elec} we only display the results for $\mu^+\to
e^+e^-e^-$. We observe that the branching ratio can 
saturate the current experimental bound for RH neutrino masses $M_N<0.1~\mathrm{TeV},1~\mathrm{TeV},~10$~TeV provided $|{\pmb{\lambda_1}}|\gtrsim 2\times 10^{-2},~10^{-1}, ~1$, respectively, very similar to the $\mu \to e \gamma$ case. The results for the inverted light neutrino mass hierarchy are again very similar and consistent with these values. As can be seen by comparing figs.~\ref{fig:radiative-lfv} and~\ref{fig:muto3elec}, with the sensitivities of the planned
future experiments~\footnote{The proposed {\it Mu3e} experiment at
  PSI aims for a sensitivity of $10^{-15}$ in its first phase and
  $10^{-16}$ in its second phase~\cite{psi}.} this process has the potential to probe considerably larger values of the RH neutrino masses (compared with $\mu\to e\gamma$), reaching RH neutrino mass scales in excess of  $\mathcal O(
10^5~{\rm GeV})$ for $|\pmb{\lambda_1}|\sim 1$. Finally we note that due to the strong $|\pmb{\lambda_1}|$ dependence, values of $|\pmb{\lambda_1}|$ below $10^{-3}$ are not expected to yield observable rates at near future experimental facilities even for RH neutrino masses of the order 100~GeV.

\subsection{$\mu-e$ conversion in nuclei}
\label{sec:mu-e-conversion}
Competitive lepton flavor violation constraints can also be obtained from searches for
$\mu-e$ conversion in nuclei. Currently the strongest bounds on
BR$_{\mu e} \equiv \Gamma_{\text{conversion}}/
\Gamma_{\text{capture}}$ were set by the SINDRUM collaboration from
experiments on titanium with BR$^{(\text{Ti})}_{\mu e} < 4.3 \times
10^{-12}$~\cite{Dohmen:1993mp} and gold target setting BR$_{\mu
  e}^{(\text{Au})} < 7 \times 10^{-13}$~\cite{Bertl:2006up}, both at
90\%CL.  The $\mu - e$ conversion bounds are expected to be further
improved in the future by several orders of magnitude. According to
proposals~\cite{Ankenbrandt:2006zu} and~\cite{P21Jparc, P20Jparc}, one
can expect a sensitivity of $10^{-16}$ or even $10^{-18}$ by the
PRISM/PRIME experiment.

To get the constraint in the $\mu-e$ channel from these experiments,
one needs to compute the relevant transition matrix elements in different nuclei. A
detailed numerical calculation has been carried out
by~\cite{Kitano:2002mt} and we use their formula in eq.~(14) to
calculate the desired conversion rates.  They receive one-loop
contributions from photonic penguins contributing to both effective
dipole ($A_R$) and vector ($g^{(u,d)}_{LV}$) couplings, as well as $Z$
penguins and $W$ box diagrams (these only contribute to
$g^{(u,d)}_{LV}$).  Using the notation of~\cite{Kitano:2002mt} we thus
have
\begin{equation}
  \label{eq:conversion-rate}
  \Gamma_{\text{conversion}} = 2 G_F^2 
  \left| A_R^* D +  (2g_{LV}^{(u)} + g_{LV}^{(d)}) V^{(p)} + (g_{LV}^{(u)} + 2 g_{LV}^{(d)}) V^{(n)} \right|^2 \,,
\end{equation}
where $G_F$ is the standard model Fermi coupling constant and $A_R$, $g_{LV}^{(u,d)}$ are found to be ($Q_{u,d} = 2/3, -1/3$):
\begin{align}
  A^*_R & =   \frac{\sqrt 2   }{8 G_F M_W^2} \frac{\alpha_W}{8\pi}G^{\mu e}_\gamma \,,\\
  g_{LV}^{(u)} &= \frac{\sqrt 2 \alpha_W^2}{8 G_F M_W^2} \left[ \left(F_Z^{\mu e}  + 4 F_{\rm Box}^{\mu3e(1)}\right) -  4 Q_u s_W^2 \left(F_Z^{\mu e} - F_\gamma^{\mu e}\right) \right]\,,\\
  g_{LV}^{(d)} &=  - \frac{\sqrt 2 \alpha_W^2}{8 G_F M_W^2} \left[ \left(F_Z^{\mu e}  + F_{\rm Box}^{\mu 3e(1)}\right) +  4 Q_d s_W^2 \left(F_Z^{\mu e} - F_\gamma^{\mu e}\right) \right]\,.
\end{align}

\begin{figure}
  \centering
  \includegraphics[width=7cm]{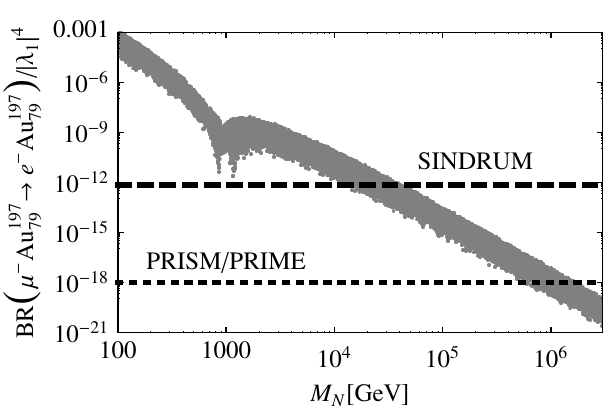}
  \includegraphics[width=7cm]{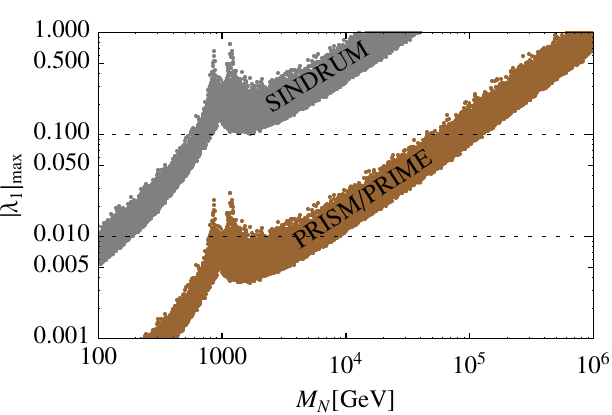}
  \caption{{\it Branching ratio BR($\mu^-\mbox{\rm Au}_{79}^{197}\to
      e^-\mbox{\rm Au}_{79}^{197}$) normalized to $|\pmb{\lambda_1}|^4$ as a function of the common RH
      neutrino mass for the light neutrino normal mass spectrum (left hand side plot). The
      upper horizontal dashed line indicates the current bound on this
      process settled by the SINDRUM experiment~\cite{Bertl:2006up},
      whereas the lower one indicates future experimental sensitivities of the PRISM/PRIME experiment~\cite{Ankenbrandt:2006zu,P21Jparc, P20Jparc}.
      The corresponding bounds on $|\pmb{\lambda_1}|$ from the present~\cite{Bertl:2006up} (upper gray band) and prospective future~\cite{Ankenbrandt:2006zu,P21Jparc, P20Jparc} (lower brown band) experimental searches are shown in the right hand side plot. 
   The widths of the bands are due to the uncertainties in the neutrino mass matrix parameters.   
      The results for the inverted neutrino spectrum are very similar and are thus not shown separately.
      }}
  \label{fig:mu-conversion}
\end{figure}

\begin{table}
  \begin{center}
  \begin{tabular}{lcccc} \hline
    Nucleus                        &$D [m_\mu^{5/2}]$  & $V^{(p)} [m_\mu^{5/2}]$ & $V^{(n)} [m_\mu^{5/2}] $ & $\Gamma_{\text{capture}}[10^6 s^{-1}]$ \\ \hline
    $\text{\text{Ti}}^{48}_{22}$    & 0.0864 &              0.0396            &             0.0468              &  2.59 \\
    $\text{\text{Au}}^{197}_{79}$ &  0.189 &              0.0974            &             0.146                &  13.07 \\ \hline
  \end{tabular}
  \end{center}
  \caption{Data taken from Tables I and VIII of~\cite{Kitano:2002mt}.}
  \label{TabTiAuData}
\end{table}
Following the same numerical procedure as in sections~\ref{sec:radiative-decays} and~\ref{sec:lto3lp} we evaluate the
resulting $\mu-e$ conversion branching ratios.  Since both Ti and Au
processes feature the same flavor structure, the differences between
them are entirely determined by the numerical factors quoted in
table~\ref{TabTiAuData}. The Ti parameters entering in the conversion
rate are on average a factor $\sim 2.5$ smaller than the ones for Au,
whereas the capture rates differ by a factor $\sim 5$. Accordingly the
difference between these branching ratios is a factor of $\sim 2$. Due
to its more stringent experimental upper bound we thus display only
the results for Au in fig.~\ref{fig:mu-conversion} for the case of the
normal light neutrino mass spectrum (the differences with the inverted
mass spectrum case are again tiny). For $M_N\sim 1$~TeV the pieces
proportional to $V^{(p),(n)}$ in 
eq.~(\ref{eq:conversion-rate}) cancel, so the $\mu-e$ conversion
rate around this RH neutrino mass value is mainly controlled by the
dipole contribution, $A_R^*D$, which is roughly two orders of magnitude
smaller. The dip in fig.~\ref{fig:mu-conversion} is due to this
cancellation.

From fig.~\ref{fig:mu-conversion} it can be seen that the current experimental bound on this 
 process imposes a constraint on the RH neutrino mass (as a function of $|\pmb{\lambda_1}|$), which is roughly a factor of 2 stronger 
compared to bounds from $\mu\to e\gamma$ and $\mu\to 3e$ (except for the region around $M_N\sim 1$~TeV, as explained above).
Furthermore,  given the expected
future sensitivities, $\mu^-\text{Au}_{79}^{197}\to
e^-\text{Au}_{79}^{197}$ (and $\mu^-\text{Ti}_{22}^{48}\to
e^-\text{Ti}_{22}^{48}$) could probe RH neutrino masses up to
$\mathcal O (10^3~\mathrm{TeV})$ , far above the values accessible in $\mu\to e\gamma$
and $\mu^-\to e^-e^+e^-$, and  thus constitutes the primary search channel for such scenarios of heavy RH neutrinos.

\section{Primordial lepton asymmetries}
\label{sec:primordial-asymm}
We now turn to the issue of primordial lepton asymmetries and the
related dynamics of the RH neutrinos.  As we have discussed, different
$R$-charge assignments allow to define two types of models of which
type B may yield sizable charged lepton flavor violating decays.  For
these effects to take place RH neutrino masses at or below the TeV  scale as well as Yukawa
couplings of order $10^{-2}$ or larger are needed. The washouts induced
by such couplings and in this mass range are so large that any lepton
asymmetry generated via the out-of-equilibrium decays of the RH
neutrinos will always yield a baryon asymmetry much smaller than
the observed one~\cite{Hinshaw:2008kr}\footnote{In models with a
  slightly broken lepton number the washout is tiny, as it is
  determined by the amount of lepton number violation
 ~\cite{Blanchet:2009kk}. In our case since lepton number is broken
  even in the $U(1)_R$ symmetric phase the washouts are dominated---as
  usual---by $N_{1,2}$ inverse decays.}.  

Either producing a baryon asymmetry consistent with the observed value
or not erasing a preexisting one via the dynamics of the RH neutrino
states (in case the RH neutrinos are still {\it light} and the
resonant condition $M_{N_2}-M_{N_1}\sim\Gamma_{N_1}$ is not satisfied)
requires small Yukawa couplings, thus rendering charged lepton flavor
violating decay branching ratios negligibly small.  The
phenomenological requirements of sizable charged lepton flavor
violating effects and the generation of a $B-L$ asymmetry (or of not
erasing a preexisting one) are therefore mutually exclusive. Since
these requirements cover non-overlapped regions in parameter space
they are from that point of view complementary.

The generation of a $B-L$ asymmetry in the type B models discussed
here follows quite closely the analysis done in ref.~\cite{Asaka:2008bj}. Thus, we do not discuss this issue here and
instead study the constraints on parameter space derived from the
condition of not erasing an assumed preexisting $B-L$ asymmetry.  Note
that in type-I seesaw models with flavor symmetries in the lepton
sector, as for example in MLFV models, the CP violating asymmetry in RH
neutrino decays vanishes in the limit of exact flavor symmetry~\cite{Bertuzzo:2009im}. However, since in type B models the MLFV
hypothesis is a consequence of the intrinsic structure of the model
this does not happen.

In order to quantify these effects from now on we focus on the
normal hierarchical light neutrino spectrum. Results for the inverted
hierarchical case resemble quite closely the ones reported here. 
We start by recalling that the washouts induced by both RH neutrino states (at
$T\sim M$) on any primordial $B-L$ asymmetry are determined by
the following set of kinetic equations:
\begin{equation}
  \label{eq:kinetic-eqs}
  \frac{d Y_{\Delta_i}}{dz}=-\frac{\kappa_i}{4}
  \sum_{j=e,\mu,\tau}C^{(\ell)}_{ij}\,Y_{\Delta_j}\,K_1(z)z^3\,.
\end{equation}
Here $Y_X=(n_X-n_{\bar X})/s$ (where $n_X$ is the number density of
particle $X$ and $s$ is the entropy density), $z=M/T$ and
$\Delta_i=B/3-L_i$ with $L_i=2\ell_i + e_i$. The function $K_1$ is the
modified Bessel function of the first type and the flavor coupling
matrix $\pmb{C^{(\ell)}}$ is determined by the chemical equilibrium
conditions imposed by the reactions that at the relevant temperature
regime ($T\sim M$) are in thermal equilibrium
\cite{Nardi:2006fx}. The parameter $\kappa_i$, that determines the
strength of the flavored washouts, is given by
\begin{equation}
  \label{eq:kappa-factor}
  \kappa_i=\frac{\tilde m_i}{m_\star}
  \quad \mbox{where}\quad 
  \tilde m_i=2\frac{v^2}{M}|\lambda_{i1}|^2\,.
\end{equation}
The factor $m_\star\simeq 1.1\times 10^{-12}$ GeV. Note that in the
basis in which the RH Majorana neutrino mass matrix is diagonal
$N_{1,2}$ couple to the lepton doublets with strength $\lambda_{i1}$,
the factor 2 in $\tilde m_i$ is due to this fact.

\begin{figure}
  \centering
  \includegraphics[width=7.4cm,height=6cm]{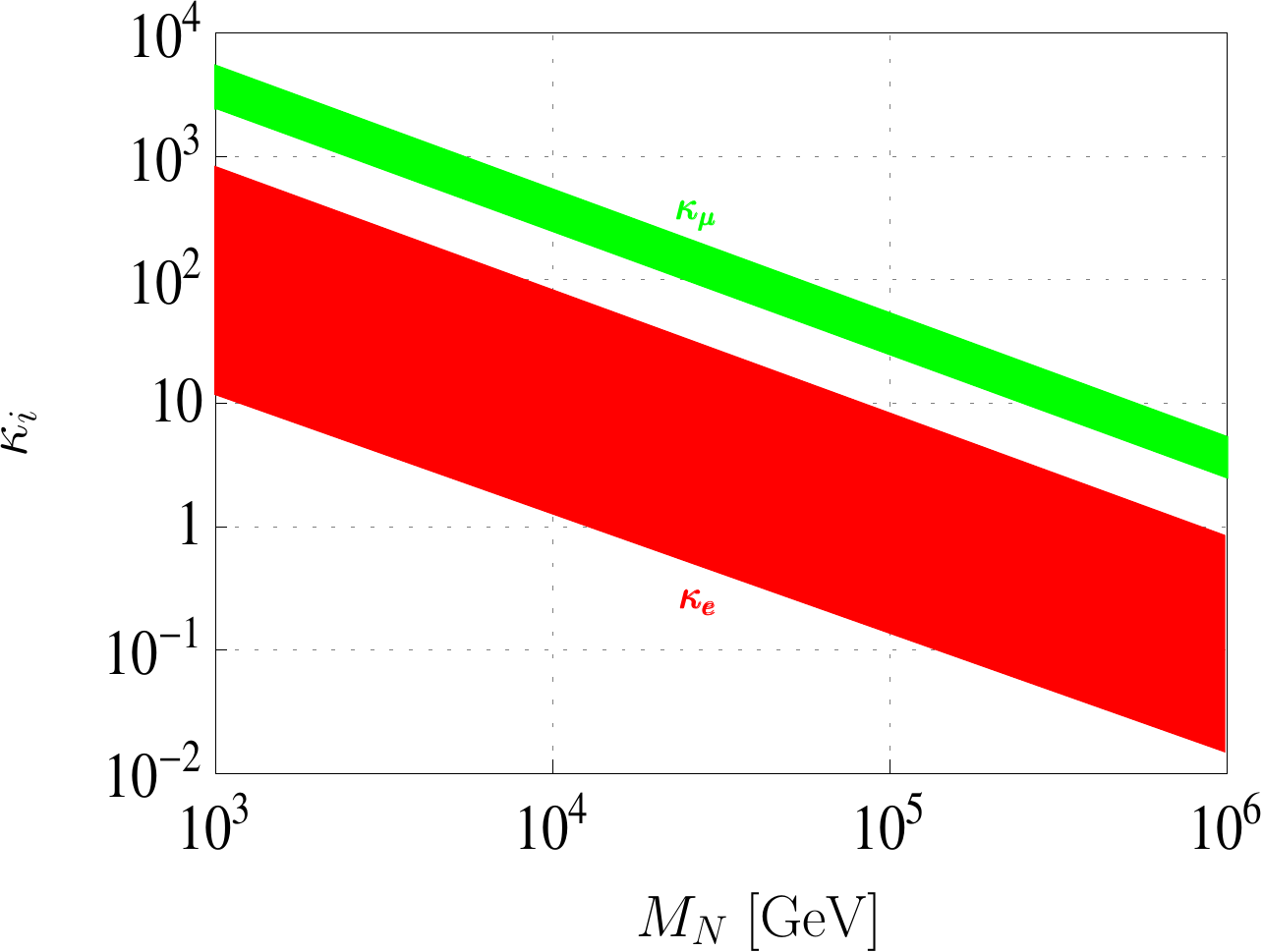}
  \includegraphics[width=7.4cm,height=6cm]{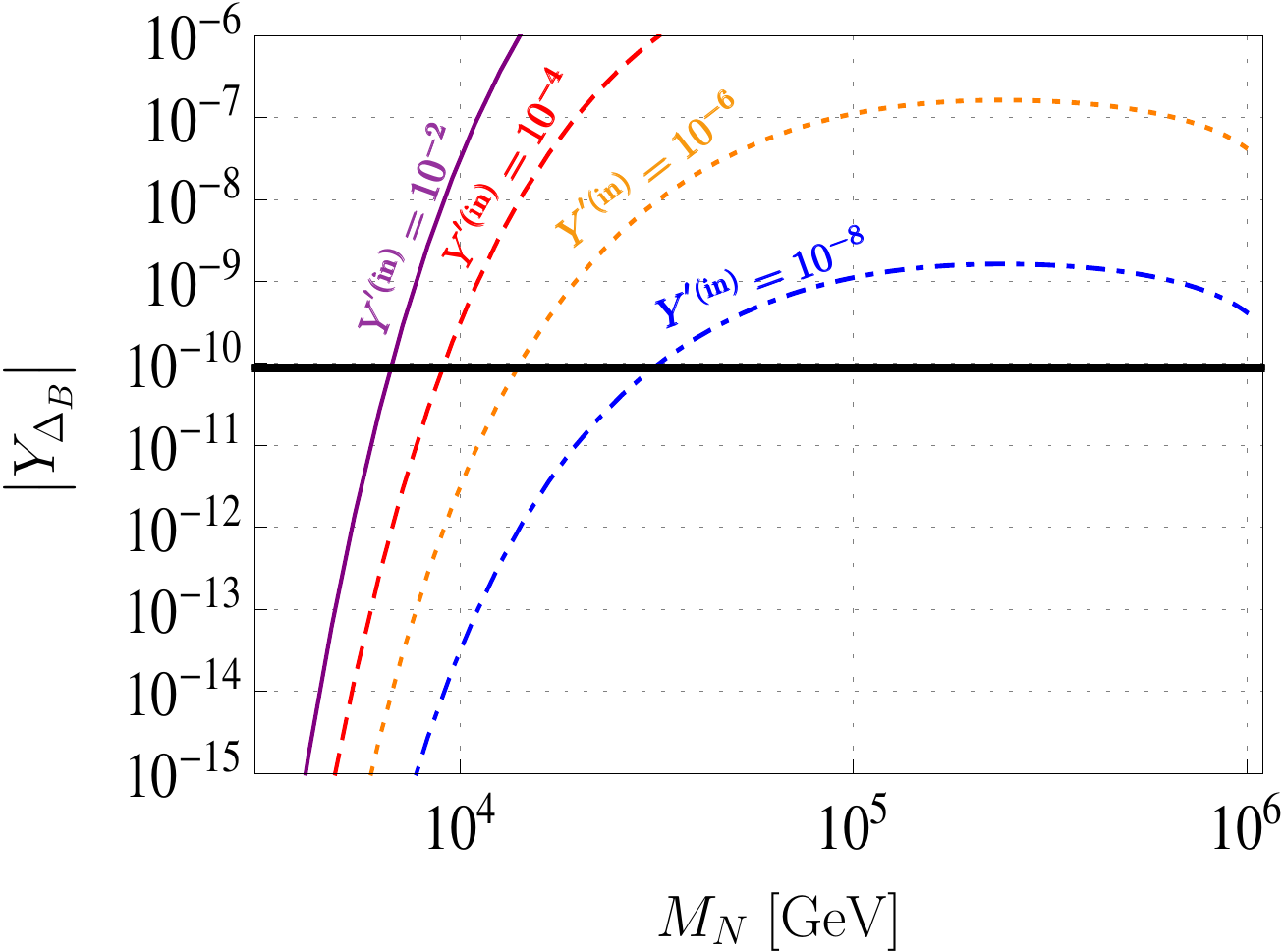}
  \caption{\it Left hand side plot: washout factors for muon and electron
    lepton flavors as a function of the common RH neutrino mass in the
    case of a normal hierarchical spectrum. Right hand side plot:
    $|Y_{\Delta_B}|$ as a function of the common RH neutrino mass for
    several values of the assumed primordial $B-L$ asymmetry. The
    solid (black) horizontal line indicates the observed value of the
    baryon asymmetry. See the text for more details.}
  \label{fig:kappa}
\end{figure}
According to the parametrization in eq.~(\ref{eq:Yukawas-normalS1}) the
$\kappa_i$ parameters can be written as
\begin{align}
  \label{eq:kappa-mix-angles}
  \kappa_i=\frac{v^2}{m_\star}\,\frac{|\pmb{\lambda_1}|^2}{M}
  |\hat \lambda_{i1}|^2
  =\frac{v^2}{m_\star}\,\frac{|\pmb{\lambda_1}|^2}{M}
  \left|
    \sqrt{1+\rho}\,U_{i3}^*
    +
    \sqrt{1-\rho}\,U_{i2}^*
  \right|^2\,.
\end{align}
Thus, after fixing low-energy observables the values of the parameters
$\kappa_i$ depend only on $M$ and
$|\pmb{\lambda_1}|$. Fig.~\ref{fig:kappa} (left hand side plot) shows
an example for the values of $\kappa_{e,\mu}$ (the $\kappa_\tau$ is
smaller than $\kappa_\mu$ by less than a factor 10) obtained by
enforcing neutrino data to lie within their $2\sigma$ experimental
ranges~\cite{Schwetz:2011zk} and fixing for concreteness
$|\pmb{\lambda_1}|=10^{-5}$. As can be seen, if the preexisting
asymmetry is sufficiently large even in the case of light RH neutrinos
a sizable asymmetry in the electron flavor could be stored.

An estimation of the $N_{1,2}$ washout effects can be easily done in
the one-flavor approximation by taking
$\pmb{C^{(\ell)}}=\pmb{\mathds{I}}$ in eq.~(\ref{eq:kinetic-eqs}). The
resulting equation can be analytically integrated yielding the
following result for the final baryon asymmetry:
\begin{equation}
  \label{eq:final-BA-nonFlavored-regime}
  Y_{\Delta_B}=\frac{12}{37}\,Y_{\Delta_{B-L}}^{(\text{in})}\,
  e^{-3\pi \kappa/8}\,.
\end{equation}
From this equation a parametric relation between the relevant
parameters can be calculated, namely
\begin{equation}
  \label{eq:lambda1overM}
  \frac{|\pmb{\lambda_1}|^2}{M}=\frac{8m_\star}{3\pi v^2}
  \log\left(
    \frac{12}{37}
    \frac{Y_{\Delta_{B-L}}^{(\text{in})}}{Y_{\Delta_B}}
  \right)\,,
\end{equation}
thus fixing $Y_{\Delta_B}$ to its central value
($Y_{\Delta_B}=8.75\times 10^{-11}$ \cite{Hinshaw:2008kr}) and taking
$Y_{\Delta_{B-L}}^{(\text{in})}\subset [10^{-8},10^{-2}]$ it turns out
that as long as $|\pmb{\lambda_1}|^2/M\subset [1,5]\times
10^{-16}\;\mbox{GeV}^{-1}$ a primordial asymmetry may always survive
the $N_{1,2}$ related washouts and yield a value consistent with the
observed one.

A precise treatment, however, requires the inclusion of flavor. In the
mass range we are interested in ($[10^3,10^6]$ GeV) all the standard
model Yukawa processes (quarks and leptons) are in thermodynamical
equilibrium~\cite{Nardi:2006fx}. Neglecting order one spectator
processes, the kinetic eqs. (\ref{eq:kinetic-eqs}) consist of
three coupled differential equations accounting for the evolution of
the $\Delta_{\tau,\mu,e}$ asymmetries. Defining the {\it asymmetry
  vector}
$\pmb{Y_{\Delta}}=(Y_{\Delta_\tau},Y_{\Delta_\mu},Y_{\Delta_e})$ the
system of coupled equations can be arranged in a single equation
\begin{equation}
  \label{eq:coupled-system-eqs}
  \frac{d}{dz}\pmb{Y_\Delta}=-\frac{v^2}{4 m_\star}
  \frac{|\pmb{\lambda_1}|^2}{M}
  \,\pmb{\tilde C^{(\ell)}}
  \;\pmb{Y_\Delta}K_1(z)z^3\,,
\end{equation}
where $\tilde C^{(\ell)}_{ij}=|\hat \lambda_{i1}|^2C^{(\ell)}_{ij}$
and the matrix $\pmb{C^{(\ell)}}$, at this stage, is given by
\cite{Nardi:2006fx}
\begin{equation}
  \label{eq:ctilde-ell}
  \pmb{C^{(\ell)}}=\frac{1}{711}
  \begin{pmatrix}
     221 & -16 & -16\\
     -16 & 221 & -16\\
     -16 & -16 & 221
  \end{pmatrix}\,.
\end{equation}
By rotating the {\it asymmetry vector} in the direction in which
$\pmb{\tilde C^{(\ell)}}$ becomes diagonal
($\pmb{Y_{\Delta}}'=\pmb{P}\,\pmb{Y_{\Delta}}$) the system of
equations can be decoupled and thus solved analytically for
$\pmb{Y_{\Delta}}'$ as in the unflavored regime:
\begin{equation}
  \label{eq:decoupled-system-eqs}
  \frac{d}{dz}\pmb{Y_\Delta}'=-
  \frac{v^2}{4m_\star}
  \frac{|\pmb{\lambda_1}|^2}{M}
  \,\pmb{\tilde C^{(\ell)}_\text{diag}}
  \;\pmb{Y_\Delta}'K_1(z)z^3
  \quad\mbox{with}\quad
  \pmb{P}\pmb{\tilde C^{(\ell)}}\pmb{P^{-1}}=
  \pmb{\tilde C^{(\ell)}_\text{diag}}\,.
\end{equation}
The solution reads
\begin{equation}
  \label{eq:solution-decoupled}
  Y_{\Delta_i}'=Y_{\Delta_i}^{'(\text{in})}\,
  e^{-3\pi \kappa \tilde c_i/8}\,,
\end{equation}
where the $\tilde c_i$'s ($i=\tau,\mu,e$) are the eigenvalues of the
matrix $\pmb{\tilde C^{(\ell)}}$. The final baryon asymmetry in this
case is therefore given by
\begin{equation}
  \label{eq:final-BA-flavored}
  Y_{\Delta_B}=\frac{12}{37}\sum_{j=\tau,\mu,e}Y_{\Delta_j}=
  \frac{12}{37}\sum_{j,i=\tau,\mu,e}\left(\pmb{P}^{-1}\right)_{ji}
  Y_{\Delta_i}^{'(\text{in})}\,e^{-3 \pi \kappa \tilde c_i/8}\,.
\end{equation}
In order to illustrate the effects of the $N_{1,2}$ related washouts
on a preexisting $B-L$ asymmetry we fix the light neutrino mixing
angles and the atmospheric and solar scales to their best fit point
values~\cite{Schwetz:2011zk}, $\delta=\pi/2$, $\phi=0$ and again
$|\pmb{\lambda_1}|=10^{-5}$. Assuming the same primordial $\Delta_i$
asymmetries in each flavor, varying them from $10^{-8}-10^{-2}$, and
using eq.~ (\ref{eq:final-BA-flavored}) we calculate the resulting
$Y_{\Delta_B}$ asymmetry. The results are displayed in
fig.~\ref{fig:kappa} (right hand side plot). It can be seen that for
the set of parameters chosen a $Y_{\Delta_B}$ in the observed range
can always be obtained.

\section{Conclusions}
\label{sec:conc}
Besides the global total lepton number $U(1)_L$ the canonical seesaw
mechanism also breaks a global $U(1)_R$ symmetry respected by the kinetic and gauge terms in the SM Lagrangian. In the context of MLFV models,
this $U(1)_R$ can be identified with global phase rotations of the charged lepton
electroweak singlets $e$ or RH neutrinos $N$. In this paper we have
explored the implications of a slightly broken $U(1)_R$ symmetry in
the context of minimal seesaw setups (with two RH neutrinos). We have shown
that depending on the $R$-charge assignments two classes of generic
models can be identified: (type $A$) models where the small breaking of
$U(1)_R$ allows to decouple the lepton number breaking scale from the
RH neutrino mass scale~\cite{Alonso:2011jd}; (type $B$)
models where the parameters that slightly break the $U(1)_R$ induce a
suppression in the light neutrino mass matrix.

We have studied the implications of these models for charged lepton flavor
violating decays. We found that in type A models the decoupling of the
RH neutrino masses from the lepton number breaking scale implies also
a suppression of the corresponding Yukawa couplings, thus leading to non-observable
charged lepton flavor violating effects. Type B models realize the
MLFV hypothesis in the sense that due to the structure of the light
neutrino mass matrix their flavor patterns are---up to normalization
factors---entirely determined by low-energy neutrino
observables. Moreover, the suppression induced by the slightly broken
$U(1)_R$ on the neutrino mass matrix allows large Yukawa couplings and
TeV RH neutrino masses, and thus potentially large flavor violating
$\mu$ processes. We have studied the $\mu\to e\gamma$, $\mu\to 3e$ and
$\mu-e$ conversion in nuclei for normal and inverted neutrino mass
spectra, finding that the three processes have branching
ratios accessible in present experiments as long as the relevant overall Yukawa normalization factor is larger than $\sim 10^{-2}, 10^{-1}, 1$ and the
RH neutrino masses are below $\sim 0.1~\mathrm{TeV}, 1~\mathrm{TeV}, 10~$TeV, respectively. For heavier RH
neutrinos $\mu\to e \gamma$ is below prospective future sensitivities while $\mu\to 3e$ and $\mu-e$ conversion in nuclei would remain observable, up to
$M_N \sim 100$~TeV and $M_N\sim 10^{3}$~TeV respectively. On the other hand in both type A and B models, RH neutrino contributions to LFV tau lepton decays are restricted below the present and near future experimental sensitivities.

Sizable $\mu$ flavor violating decays require large Yukawa couplings
and {\it light} RH neutrinos. These values imply large RH neutrino
inverse decay effects, that render the dynamics of these states
incompatible with either the generation of a $B-L$ asymmetry
(consistent with the observed $B$ asymmetry) or with the preservation
of a preexisting one.  Accordingly, sizable lepton flavor processes
and small RH neutrino inverse decay effects are phenomenological
requirements that cover non-overlapping regions of parameter space,
from that point of view the analysis of both of them turns out to be
complementary. In the {\it low mass} range ($M\lesssim 10^6$ GeV),
instead of studying the generation of a $B-L$ asymmetry via resonant
leptogenesis, we have considered the influence of the RH neutrino dynamics on
a primordial $B-L$ asymmetry. We have demonstrated that  a preexisting asymmetry yielding
the observed $B$ asymmetry can survive the RH neutrino related
washouts provided the overall Yukawa
coupling normalization is below $\sim 10^{-5}$ .

\section*{Acknowledgments}
We want to thank G. Isidori and E. Nardi and M. Hirsch for useful
comments and remarks. DAS is supported by a belgian FNRS
fellowship. The work of JFK was supported in part by the Slovenian
Research Agency. DAS and AD want to thank the Josef Stefan Institute
for the kind hospitality during the completion of this work.

\appendix

\section{Formulas for $l_i\to l_j \gamma$ and $l_i^-\to l_j^- l_j^+
  l_j^-$ processes}
\label{sec:formulas-for-lfvprocesses}
In this appendix we summarize the formulas we use for the calculation
of the charged lepton flavor violating decays discussed in
sections~\ref{sec:radiative-decays} and~\ref{sec:lto3lp}. The
results presented here were extracted from ref.~\cite{Ilakovac:1994kj} and adapted to our notation. In what follows
the parameters $r_a$'s are defined according to
$r_a=M_W^2/M_{N_a}^2$. 

The process $l_i^-\to l_j^- l_j^+ l_j^-$ is determined by $\gamma$,
and $Z$ penguins and box diagrams (for the full set of Feynman
diagrams see ref.~\cite{Ilakovac:1994kj}). The $\gamma$ penguin
contribution can be split in two pieces corresponding to the photon
being either on-shell or off-shell. For the on-shell piece, the one
that determines the $l_i\to l_j\gamma$ process, we have
\begin{align}
  \label{eq:gfac}
  G_\gamma^{l_i l_j}&=\frac{2}{g^2}
  \left(
  \pmb{\lambda}\cdot\pmb{G_\gamma}\cdot\pmb{\lambda}^\dagger
  \right)_{ij}\,,\\
  G_{\gamma}(r_a)&=\frac{r_a}{4(1-r_a)^4}
  \left(2+3r_a-6r_a^2+r_a^3+6r_a \log r_a\right)\,,
\end{align}
whereas for the off-shell photon piece
\begin{align}
  \label{eq:ffac}
  F_\gamma^{l_i l_j}&=\frac{2}{g^2}
  \left(
    \pmb{\lambda}\cdot\pmb{F_\gamma}\cdot\pmb{\lambda}^\dagger
  \right)_{ij}\,,\\
  F_{\gamma}(r_a)&=-\frac{r_a}{12(1-r_a)^4}
  \left[7-8r_a-11r_a^2+12r_a^3-(2- 20r_a+24r_a^2) \log r_a\right]\,.
\end{align}
The $Z$ penguin contribution can be split in two parts, namely
\begin{equation}
  \label{eq:Z-form-factor-complete}
  F_Z^{l_i l_j}=F_Z^{l_i l_j(1)}+F_Z^{l_i l_j(2)}\,,
\end{equation}
where the first piece can be written as
\begin{align}
  \label{eq:zffacsnf}
  F_Z^{l_i l_j(1)}&=\frac{2}{g^2}\left[\pmb{\lambda}\cdot\left(\pmb{\hat F_Z}
      + \pmb{\hat G_Z^{(1)}}\right)\cdot\pmb{\lambda}^\dagger\right]_{ij}\,,\\
  F_{Z}(r_a)&=\frac{5r_a}{2(1-r_a)^2}\left(1 - r_a + \log r_a\right)\,,\\
  G_{Z}^{(1)}(r_a)&=-\frac{r_a}{1-r_a}\log r_a\,,
\end{align}
while the second contribution according to
\begin{align}
  \label{eq:dnfdiag}
  F_Z^{l_i l_j(2)}&=\frac{4}{g^4}
  \left[\pmb{\lambda}\cdot
    \left(
      \pmb{\tilde G_Z^{(2)}}
      + \pmb{\tilde G_Z^{(3)}}
      + \pmb{\tilde G_Z^{(4)}}
      + \pmb{\tilde H_Z}
    \right)
    \cdot\pmb{\lambda}^\dagger
  \right]_{ij}\,,\\
  \tilde G_Z^{(A)}(r_a,r_b)&=
  (\pmb{\lambda}^\dagger\cdot\pmb{\lambda})_{ab}G_Z^{(A)}(r_a,r_b)
  \quad\mbox{with}\quad A=2,3,4\,,\\
  G_Z^{(2)}(r_a,r_b)&=-\frac{r_ar_b}{2(r_a-r_b)}
  \left(
    \frac{1-r_b}{1-r_a}\log r_a
    -
    \frac{1-r_a}{1-r_b}\log r_b
  \right)\,,\\
  G_Z^{(3)}(r_a,r_b)&=\frac{r_a r_b}{2(1-r_a)}\log r_a\,,\\
  G_Z^{(4)}(r_a,r_b)&=\frac{r_ar_b}{2(1-r_b)}\log r_b\,,\\
  \tilde H_Z(r_a,r_b)&=
  \left(\pmb{\lambda}^T\cdot \pmb{\lambda}^*\right)_{ab}H_Z(r_a,r_b)\,,\\
  H_Z(r_a,r_b)&=-\frac{\sqrt{r_ar_b}}{4(r_a-r_b)}
  \left[
    \frac{r_b(1 - 4 r_a)}{1 - r_a}\log r_a
    -
    \frac{r_a(1 - 4 r_b)}{1 - r_b}\log r_b
  \right]\,.
\end{align}
Note that due to the constraint implied by the $SU(3)_{\ell+N}$ flavor
symmetry the off-diagonal elements of the matrices $\tilde
G_Z^{(A)}(r_a,r_b)$ and $\tilde H_Z^{(A)}(r_a,r_b)$ vanish.

The box diagram contributions can be split in three parts as follows
\begin{equation}
  \label{eq:box-ffac}
  F^{l_i 3l_j}_\text{Box}=\sum_{A=1,2,3}F^{l_i 3l_j(A)}_\text{Box}\,,
\end{equation}
For the first part we have
\begin{align}
  \label{eq:boxes1}
  F^{l_i 3l_j(1)}_\text{Box}&=\frac{2}{g^2}
  \left[
    \pmb{\lambda}\cdot\pmb{\hat F_\text{Box}^{(1)}}\cdot
    \pmb{\lambda}^\dagger
  \right]_{ij}\,,\\
  F_\text{Box}^{(1)}(r_a)&=-\frac{2r_a}{(1-r_a)^2}
  \left(
    1 - r_a + r_a\log r_a
  \right)\,.
\end{align}
For the second is given by
\begin{align}
  \label{eq:boxes2}
  F^{l_i 3l_j(2)}_\text{Box}(j)&=\frac{4}{g^4}
  \left[
    \pmb{\lambda}\cdot
    \left(
      \pmb{\tilde F_\text{Box}^{(2)}}(j)
      +
      \pmb{\tilde F_\text{Box}^{(3)}}(j)
    \right)  
      \cdot
    \pmb{\lambda}^\dagger
  \right]_{ij}\,,\\
  \tilde F_\text{Box}^{(A)}(r_a,r_b)(j)&=\lambda^*_{ja}\;
  F_\text{Box}^{(A)}(r_a,r_b)\;\lambda_{jb}\quad\mbox{with}\quad A=2,3\,,\\
  F_\text{Box}^{(2)}(r_a,r_b)&=\frac{r_ar_b}{4(r_a-r_b)}
  \left[
    \frac{1-4r_a(2-r_b)}{(1-r_a)^2}\log r_a
    -
    \frac{1-4r_b(2-r_a)}{(1-r_b)^2}\log r_b
    \right .
    \nonumber\\
    &\left .
      -
      \frac{r_a-r_b}{(1-r_a)(1-r_b)}
      (7 - 4r_ar_b)
    \right]\,,\\
    F_\text{Box}^{(3)}(r_a,r_b)&=2r_ar_b
    \left[
      \frac{r_b}{(1-r_b)^2}(1-r_b+\log r_b)
    +
    \frac{1}{(1-r_a)^2}(1-r_a+r_a\log r_a)
  \right]\,,
\end{align}
where in $\tilde F_\text{Box}^{(A)}(r_a,r_b)(j)$ no summation over the
indices $a, b$ is performed. Finally, the third term in (\ref{eq:box-ffac})
can be written as
\begin{align}
  F^{l_i 3l_j(3)}_\text{Box}(j)&=\frac{4}{g^4}
  \left[
    \pmb{\lambda}
    \cdot
    \pmb{\tilde G_\text{Box}}(j)
    \cdot
    \pmb{\lambda}^\dagger
  \right]_{ij}\,,\\
  \tilde G_\text{Box}(r_a,r_b)(j)&=\lambda_{ja}\;G_\text{Box}(r_a,r_b)
  \;\lambda_{jb}^*\,,\\
  G_\text{Box}(r_a,r_b)&=-\frac{\sqrt{r_ar_b}}{r_a-r_b}
  \left[
    \frac{r_a\left[1-2r_b(1-2r_a)\right]}{(1-r_a)^2}\log r_a
    -
    \frac{r_b\left[1-2r_a(1-2r_b)\right]}{(1-r_b)^2}\log r_b
  \right .\nonumber\\
  &\left .
   + \frac{(r_a-r_b)}{(1-r_a)(1-r_b)}(1+2r_a r_b)
  \right]\,,
\end{align}
where, again, in $\tilde G_\text{Box}^{(A)}(r_a,r_b)(j)$ no summation over
the indices $a$ and $b$ is performed.

\end{document}